\documentclass[]{raa}
\usepackage{graphicx,times}
\usepackage{natbib}

\begin{document}

\title{Variations in the X-ray eclipse transitions of Cen X-3}

\volnopage{ {\bf 20xx} Vol.\ {\bf 9} No. {\bf XX}, 000--000}
   \setcounter{page}{1}

\author{Jincy Devasia \inst{1,2}
\and Biswajit Paul \inst{2} \and  Marykutty James \inst{1,2} \and
Kavila Indulekha \inst{1}
 }

\institute{School of Pure \& Applied Physics, Mahatma Gandhi University, Kottayam-686560, Kerala, India.; {\it jincydevasia@yahoo.com} \and Raman Research Institute, Sadashivnagar, C. V. Raman Avenue, Bangalore 560080, India.\\
\vs \no
   {\small Received [year] [month] [day]; accepted [year] [month] [day] }
}

\abstract{We report here an investigation of the X-ray eclipse transitions of the high mass X-ray binary pulsar Cen X-3 in different intensity states. Long term light curve of Cen X-3 obtained with {\sl RXTE}-ASM spanning for more than 5000 days shows strong aperiodic flux variations with low and high states. We have investigated the eclipse transitions of Cen X-3 in different intensity states with data obtained from pointed observations with the more sensitive instruments on board {\sl ASCA}, {\sl BeppoSAX}, {\sl XMM}-Newton, {\sl Chandra} and {\sl RXTE}. We found a very clear trend of sharp eclipse transitions in the high state and longer transitions in the low state. This is a confirmation of this feature first observed with the {\sl RXTE}-ASM but now with much better clarity. From the light curves obtained from several missions, it is seen that the eclipse egress in the low state starts earlier by an orbital phase of 0.02 indicating that the observed X-rays originate from a much larger region. We have also performed spectral analysis of the post-eclipse part of each observations. From {\sl BeppoSAX} observations, the out-of-eclipse X-ray fluxes is found to differ by a factor of $\sim$ 26 during the high and low intensity states while the eclipse count rates differ by a factor of only $\sim$ 4.7. This indicates that in the low state, there is an additional scattering medium which scatters some of the source photons towards the observer even when the neutron star is completely eclipsed. We could also resolve the three iron line components using {\sl XMM}-Newton observation in the low state. By comparing the iron line equivalent width during the high and low states, it is seen that the width of iron line is relatively large during the low state which supports the fact that significant reprocessing and scattering of X-rays takes place in the low state. 
}

 \authorrunning{J. Devasia, B. Paul, M. James \& K. Indulekha}            %author_head in even pages
   \titlerunning{ Variations in the X-ray eclipse transitions of Cen X-3}  % title_head in odd pages
   \maketitle

\section{Introduction}
Cen X-3 was discovered in May 1967 by a rocket-borne detector (Chodil et al. 1967) meant for measurements of cosmic X-ray sources. Later, observations with {\sl Uhuru} satellite revealed its eclipsing binary nature and coherent pulsations (Giacconi et al. 1971; Schreier et al. 1972). The X-ray source is a persistent source now known to be an accretion-powered neutron star orbiting a highly reddened O-type supergiant companion V779 Cen (Krzeminski,
1974) every 2.08 days while rotating itself with a pulse period of $\sim$ 4.84 s. Cen X-3 is one among a few X-ray pulsars known in which the observation of X-ray eclipses has permitted the determination of both orbital and stellar parameters. The mass accretion rate onto the compact object from its companion star is usually determined from its X-ray luminosity. It is also often assumed that the averaged observed X-ray flux represents the averaged intrinsic X-ray luminosity. Although a strong stellar wind emanates from the companion star, the luminosity of the X-ray source suggests that the predominant mode of accretion is via a disk, fed by a Roche-lobe overflow which is possible only if the orbital size is small enough (Tjemkes et al. 1986). The QPOs discovered at $\sim$ 40 mHz and also the secular spin-up trend exhibited by Cen X-3 strongly supports the existence of an accretion disk around the neutron star (Takeshima et al. 1991; Tsunemi et al. 1996). Studies related to QPOs in Cen X-3 concluded that the QPO frequency has no dependence on X-ray intensity indicating that the observed luminosity of Cen X-3 may not be directly related to the mass accretion rate (Raichur $\&$ Paul 2008b). Long-term observations of Cen X-3 revealed that the X-ray intensity varies aperiodically between low and high states (Schreier et al. 1976). The average X-ray flux has been observed to vary by a factor of $\sim$ 40 between the extended low and high states. There exist two distinct spectral states in Cen X-3 during high state, and in each outburst, the source follows either of the two spectral states (Paul et al. 2005).  \\

 The intensity of Cen X-3 has been monitored for the past 14 years using All Sky Monitor (ASM) on board {\sl Rossi X-ray Timing Explorer} (RXTE). The X-ray flux is found to oscillate aperiodically between high, intermediate and low states. Recent studies regarding the long-term flux variations in Cen X-3 revealed that the different flux states of Cen X-3 are due to varying degree of obscuration of the compact object by the accretion disk (Raichur \& Paul 2008a). Using {\sl RXTE}-ASM data, the eclipse ingress and egress is found to be sharp in the high state while in the low state the light curve shows a smooth flux variation with orbital phase (Raichur \& Paul 2008a). This characteristics in the low state has been suggested to be due to the absorption of the X-rays from the neutron star by the accretion disk and the scattering of X-rays by the stellar wind of the companion star. From a study of the {\sl RXTE}-ASM light curve of Cen X-3, it was found that the duration of the eclipse egress and ingress is much larger in the low intensity state indicating that in the low state, a dominant fraction of the observed X-rays come from a large scattering region. However individual X-ray eclipses of Cen X-3 are not detectable with the {\sl RXTE}-ASM, even in the high state. In this study, the intensity dependence of the orbital modulation was derived by averaging the data acquired over several hundred orbital periods. In the present work, we have used observations taken by various missions with higher sensitivities in order to have a clear idea about the eclipse transition of the system in its different states which will help to scrutinize the proposed scenario for the long term intensity variations.

\section{Observations \& Analysis}

We have analysed archived data from several observations of Cen X-3 with {\sl RXTE}-PCA, {\sl ASCA}-GIS, {\sl BeppoSAX}-MECS, {\sl XMM} EPIC-PN and {\sl Chandra}-ACIS. The observations that include eclipse ingress or egress are useful for the present study and are reported here. In Table 1 we have listed the log of observations.

 {\sl RXTE}-ASM consists of three wide-angle shadow cameras equipped with proportional counters with a total collecting area of 90 square cm. It is sensitive in the 2-10 keV energy band monitoring the appearance of transients and records long-term intensity variations seen in bright X-ray sources (Levine et al. 1996) providing 80\% sky coverage for every satellite orbit ($\sim$95 min). Cen X-3 has been continuously observing by {\sl RXTE}-ASM since 1996.

 The large area Proportional Counter Array (PCA) on board {\sl RXTE} consists of five co-aligned gas-filled Proportional Counter Units (PCUs) that are sensitive to X-ray photons in the energy range 2-60 keV with a total effective area of 6500 cm$^2$ (Jahoda et al. 1996). Its energy resolution is 18$\%$ at 6 keV having larger collecting area compared to other imaging X-ray missions. 

 {\sl BeppoSAX}  Medium Energy Concentrator Spectrometer (MECS) consists of three grazing incidence telescopes with imaging gas scintillation proportional counters in their focal planes sensitive in the energy band of 1.3-10 keV (Boella et al. 1997). We have used data from two observations using MECS detectors consisting of a low and high state for the present analysis. 

 The {\sl ASCA} satellite (Tanaka et al. 1994) carried two Solid State Imaging Spectrometers (SIS0 and SIS1) and two Gas Imaging Spectrometers (GIS2 and GIS3), each located at the focus of four identical grazing-incidence X-ray telescopes (XRT). The {\sl ASCA} detectors are sensitive to resolve narrow emission lines in energy spectrum. 

 The {\sl XMM-Newton} Observatory (Jansen et al. 2001) comprises three 1500 cm$^2$ effective area X-ray telescopes, each with a European Photon Imaging Camera (EPIC) at the focus. Two of the EPIC imaging spectrometers use MOS CCDs, and one uses pn CCDs (Struder et al. 2001). We have used data from EPIC-PN when PN operated in a small window mode using the medium filter. 
 
 {\sl Chandra} observed Cen X-3 using the High Energy Transmission Gratings (HETGS) and the spectroscopy array of the Advanced CCD Imaging Spectrometer (ACIS-S). The {\sl Chandra} X-ray Observatory has a high resolution X-ray mirror, two imaging detectors (HRC and ACIS), and two sets of transmission gratings (LETG and HETG). ACIS is an array of ten 1024 X 1024 pixel CCDs having large detection efficiency (10$-$90$ \%$) and moderate energy resolution (10-50) over a 0.2-10.0 keV passband performing simultaneous imaging and spectroscopy.

{
\begin{table}
\caption{Log of Observations of Cen X-3}
\begin{center}
\begin{tabular}{l c c c c c}
\hline\noalign{\smallskip}
Telescope&Year&Obs Ids& Total exposure (ks) &State\\
\hline\noalign{\smallskip}
{\sl RXTE}-PCA   &1997&P20104&238&high\\
{\sl SAX}-MECS &1997&20373001&33&high\\
{\sl ASCA}-GIS    &1993&40006000&39&low\\
{\sl SAX}-MECS &2000&20892002&43&low\\
{\sl XMM} EPIC-PN &2001&0111010101&67&low\\
{\sl Chandra}-ACIS &2007&7511&39&low\\
\hline
\end{tabular}
\end{center}
\end{table}
}

\subsection{Long-term Observations of Cen X-3}

 Long-term light curve of Cen X-3 has been obtained from ASM on-board {\sl RXTE}. In Figure 1 we have shown the {\sl RXTE}-ASM long-term light curve of Cen X-3 binned with its orbital period of $\sim$ 2.0871 days. The light curve shows high and low intensity phases at arbitrary intervals. In the low intensity state, it is clearly seen that the ASM count rate is below 2 counts s$^{-1}$ while in the high state it is often more than 18 counts s$^{-1}$. The arrow marks in different panels of Figure 1 indicate the times during which the pointed observations with other observatories have been carried out.

\begin{figure}[htbp]
\centering
\includegraphics[width=3 cm, height=13 cm, angle =-90]{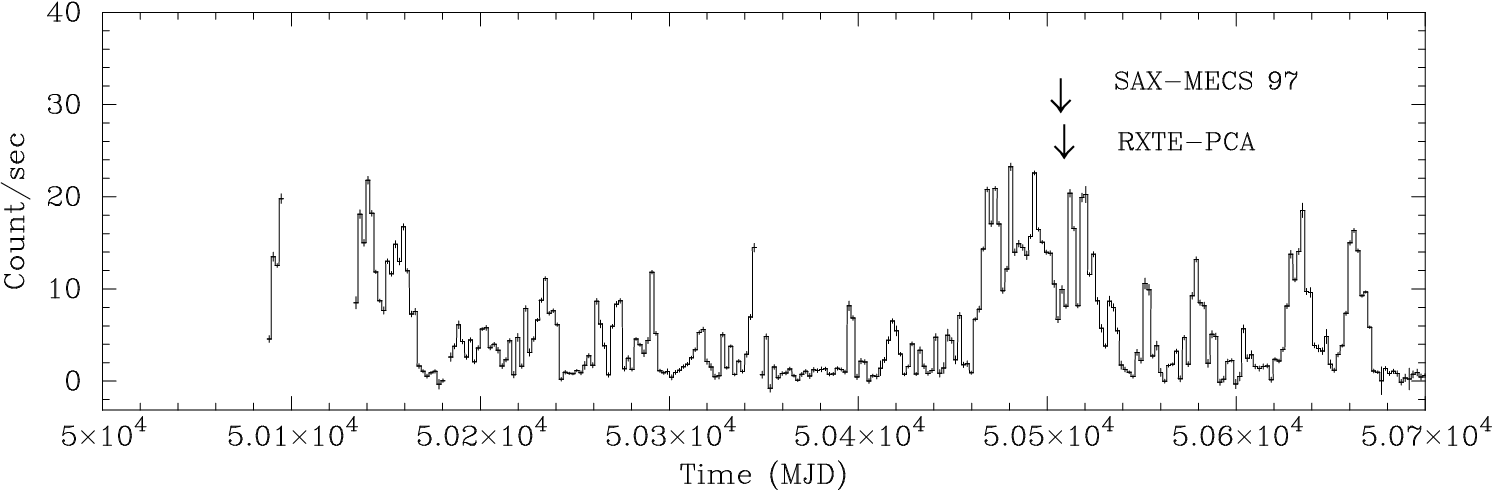}
\includegraphics[width=3 cm, height=13 cm, angle =-90]{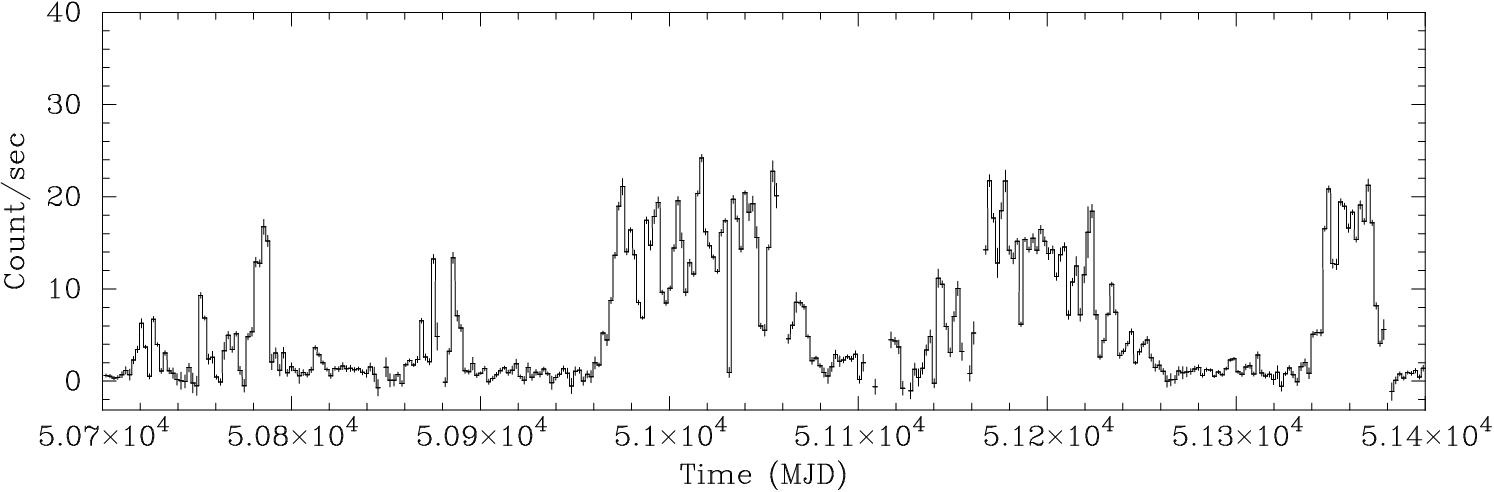}
\includegraphics[width=3 cm, height=13 cm, angle =-90]{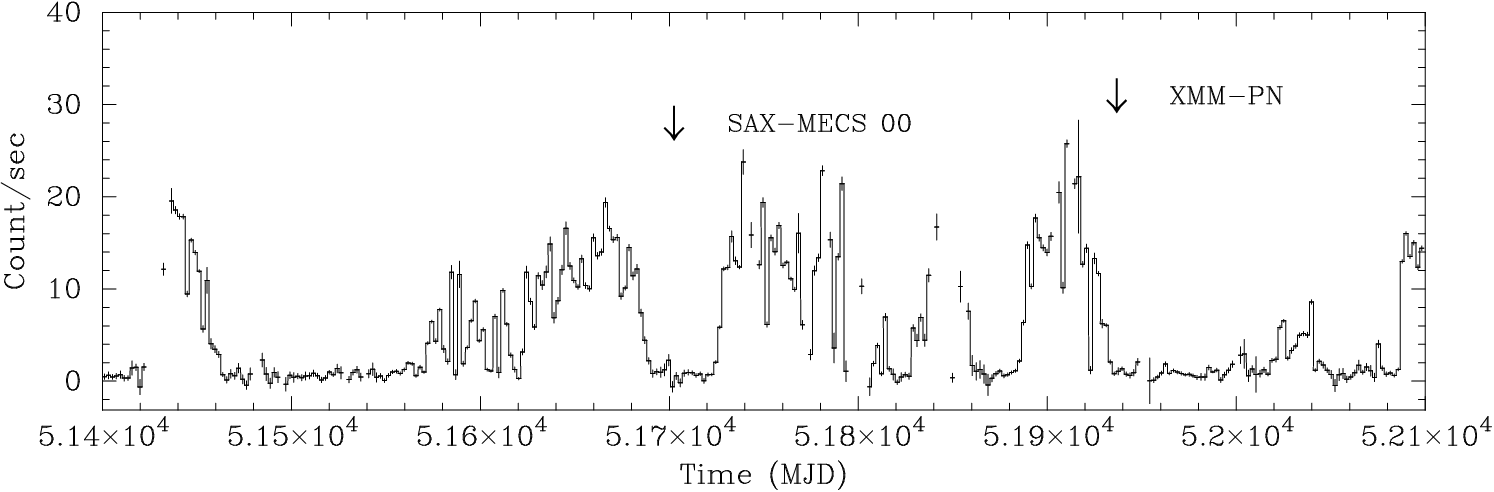}
\includegraphics[width=3 cm, height=13 cm, angle =-90]{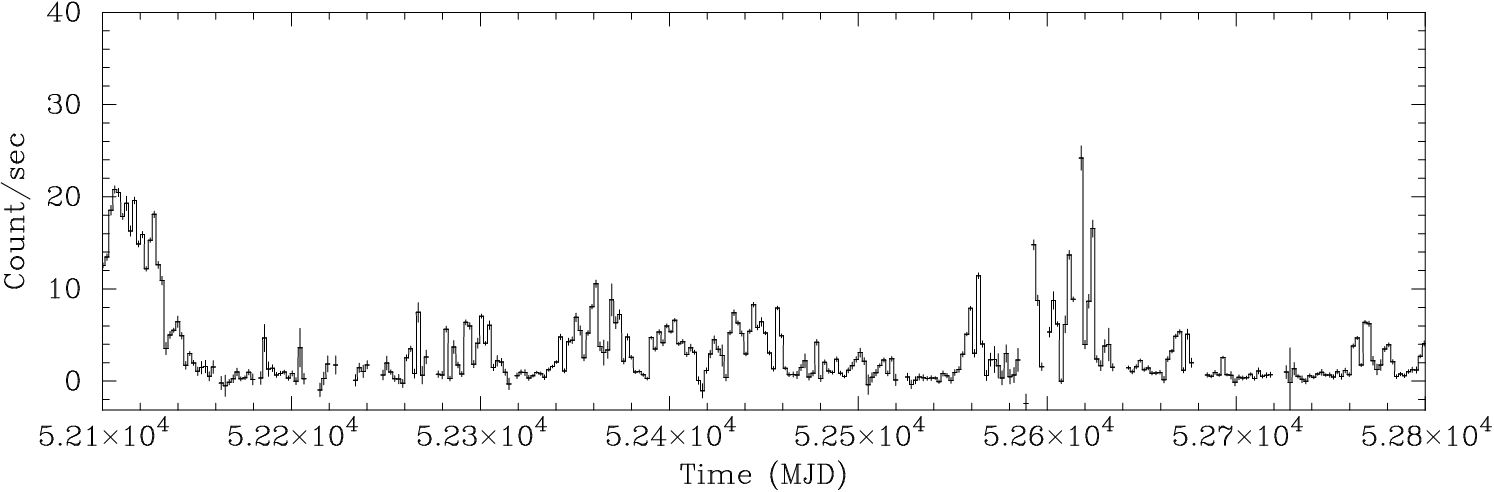}
\includegraphics[width=3 cm, height=13 cm, angle =-90]{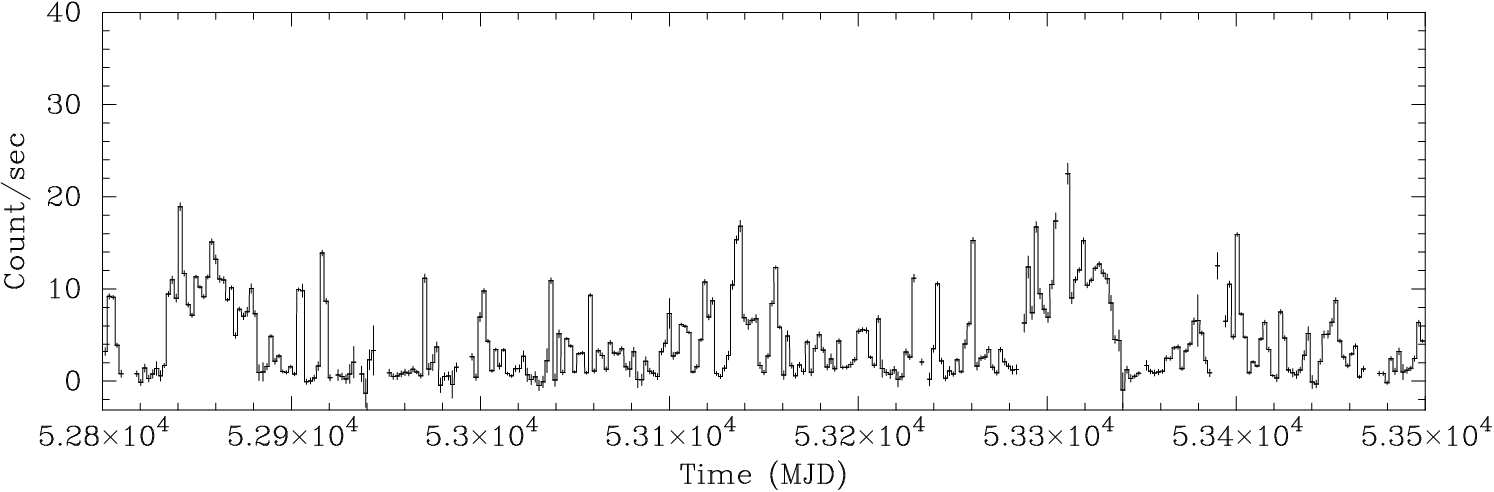}
\includegraphics[width=3 cm, height=13 cm, angle =-90]{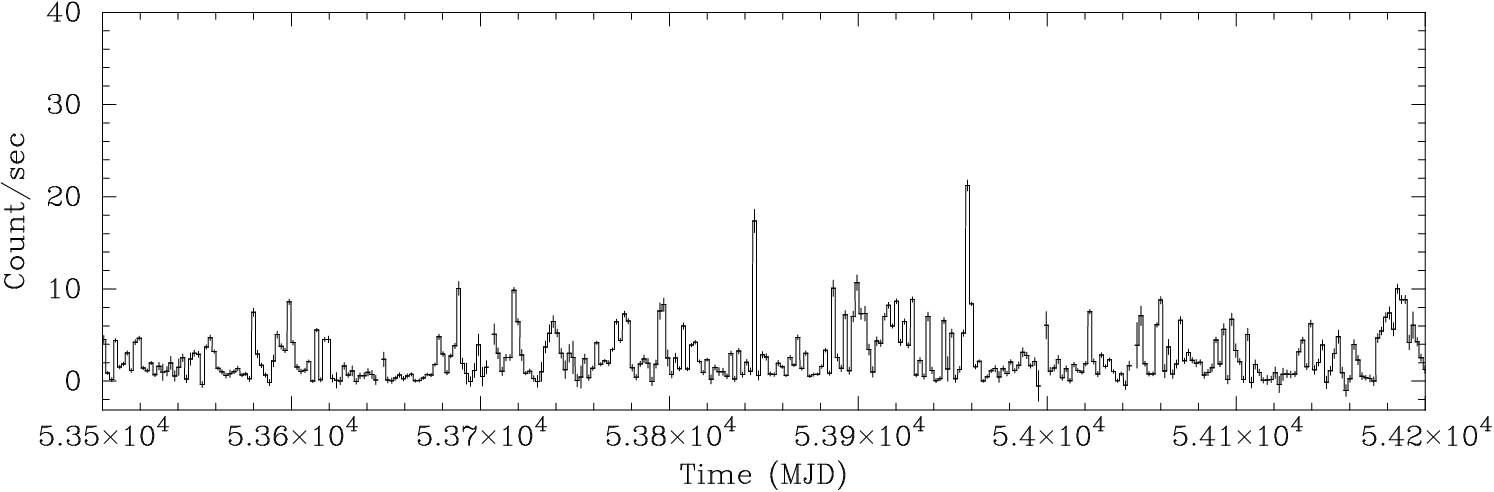}
\includegraphics[width=3 cm, height=13 cm, angle =-90]{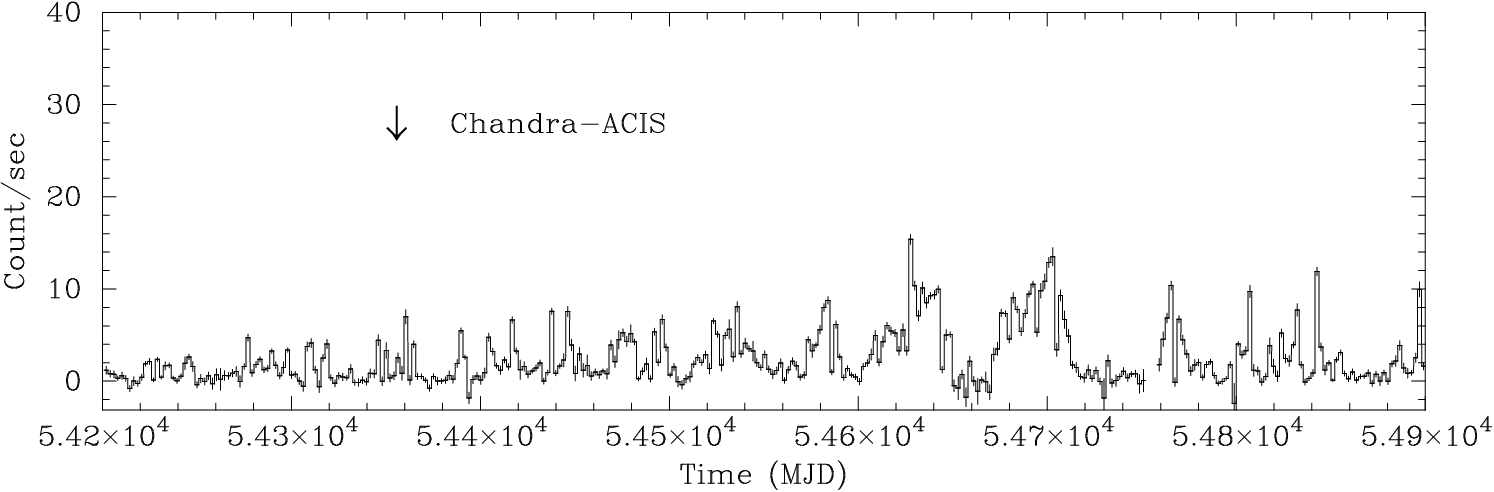}
\caption{The RXTE-ASM light curve binned with orbital period (2.08 days) of Cen X-3 is shown here. The arrow marks indicate the times during which the pointed observations were carried out.}
\end{figure}

Except the {\sl ASCA} observation, which occurred before the beginning of the {\sl RXTE}-ASM light curve, observation period for all other pointed observations are shown in the ASM light curve. Figure 1 indicates that the {\sl RXTE}-PCA observation in 1997 and the {\sl BeppoSAX} observation in
1997 were carried out when the source was in its high intensity state with an average count rate of 10 counts s$^{-1}$ and 8 counts s$^{-1}$ respectively
in the {\sl RXTE}-ASM detectors. The time period during which the {\sl XMM Newton}, {\sl Chandra} and the {\sl BeppoSAX} observation of 2000
occurred are also shown in Figure 1. During the later observations the source was in relatively low intensity state with average ASM count rate of
approximately 1 counts s$^{-1}$, 0.8 counts s$^{-1}$ and 0.6 counts s$^{-1}$ respectively. From the ASM light curve, it is noticed that the overall intensity of Cen X-3 decreased in the later years, the high state are less frequent and of shorter duration.

\subsection{Short-term Observations of Cen X-3}

{We have performed pulse folding and $\chi ^2$ maximization technique on the ASM light curve of Cen X-3 and obtained the best orbital period of 2.087081 days. This is compatible with the values reported in Paul et al. (2005) and Raichur \& Paul (2010). Using this period and the ephemeris given in Raichur \& Paul (2010), we folded the light curves obtained from several observatories and these are shown in terms of the orbital phase in Figure 2. The corresponding ASM counts for the different observations are given in each panel. From the figure, it is clearly evident that the shape of the eclipse is very strongly dependent on the overall X-ray intensity. For the spectral analysis, we have generated energy spectra from the post-egress parts of several observations. For all but the {\sl RXTE}-PCA observation, source centered and source-free regions were selected within the field of view of the instruments for extracting the source and the respective background spectra. Spectral fitting for the energy spectra created from different instruments was performed using the XSPEC v12 data analysis package.\\

 The energy spectrum of Cen X-3 is generally modeled by a power-law modified by photoelectric absorption due to the intervening interstellar matter, a black body, high-energy cutoff and a Gaussian component for iron line emission. The power-law and Gaussian components represent non-thermal emission and iron line emission respectively, which is typical of X-ray pulsars (e.g., White et al. 1983). The black body component represents the reprocessing of hard X-rays from the neutron star by the inner region of the accretion disk (Hickox et al. 2004).
The model spectrum is given by
\begin{equation}
I(E) ={ {\rm exp}[- \sigma (E) N_{\rm H}] 
       \times \{ I_{\rm P} E^{- \Gamma} + BB(E) + G(E) \} } \\
      {\rm \; \; \; photons \; cm^{-2} \; s^{-1} \; keV^{-1}} 
\end{equation}
where $E$ is the photon energy in keV; $\sigma$(E) and $N_{\rm H}$ are the photoelectric cross section and the hydrogen nuclei column density of the intervening matter; $I_{\rm P}$ and $\Gamma$ are the norm and the power-law index of the power-law component. $BB(E)$ is the black body function of temperature $kT$; $G(E)$ is the Gaussian function for the iron line emission with its center at $E_{\rm G}$, caused by fluorescence of relatively cold matter in the neutron star magnetosphere (White, Swank \& Holt, 1983). 
We used three different model spectra for different observations:
(1) the model spectrum consisting of an absorbed power-law component and Gaussian component 
(2) a model spectrum consisting of an absorbed power-law component plus blackbody and a Gaussian component
(3) the same model spectrum of (1) with a high-energy cutoff for the power-law component.
For the case (3), the model spectrum is given by
\begin{equation}
I(E) ={ {\rm exp}[- \sigma (E) N_{\rm H}]
       \times \{ I_{\rm P} E^{- \Gamma} f_{hi}(E) +  G(E) \} } \\
      {\rm \; \; \; photons \; cm^{-2} \; s^{-1} \; keV^{-1}}
\end{equation}
where
$$ f_{hi}(E) = \cases{1,&                            $E < E_c$              \cr
                      exp(- {(E - E_c) \over E_f}),& $E \geq E_c$,  \quad  \cr}
$$
represents the high-energy cutoff for the power-law component.\\

Cen X-3 was observed by {\sl RXTE}-PCA from February 28 to March 03, 1997 with a total exposure time of 238 ks covering two orbital periods. The light curves and spectra were extracted from Standard2 mode data with 16 s bin size using the tool {\sl saextrct}. The 2-60 keV light curve folded with the orbital period is shown in Figure 2 in terms of orbital phase.} The light curve shows a sharp ingress and egress with an average count rate of about 4200 counts s$^{-1}$ in the post egress part. The entire observation covers two eclipses of the X-ray source. During the PCA observation, the ASM light curve shows the source to be in a high state with an average ASM count rate of 10 counts s$^{-1}$. Background data files were generated with the tool {\sl runpcabackest} using appropriate background models for bright source provided by the PCA Calibration team. The response matrices were created using the tool {\sl pcarsp}. Model spectra were convolved with the detector response matrix and best fitted in the energy range 2.5-20.0 keV with a power-law along with a high-energy cutoff and a Gaussian line.   \\

Two observations of Cen X-3 were made with the {\sl BeppoSAX} LECS and MECS detectors during 1997 February 27-28 and 2000 June 06-08 having a useful exposure of 33 ks and 43 ks respectively. We have used combined source counts from the MECS detectors (MECS 1+2+3 in the 1997 observation and MECS 2+3 in the 2000 observation) for the present work. The light curves were extracted from circular regions of radius 6{\arcmin} around the source with time resolution of 16 s. Observation made in 2000 includes both the eclipse ingress and the egress whereas 1997 observation includes only the eclipse egress. The 1997 light curve also shows a sharp eclipse egress as is observed in the {\sl RXTE} light curve. The 1997 and 2000 light curves in terms of orbital phase are shown in Figure 2. During the 2000 observation, the eclipse has a shallow ingress and egress and the source was found to be in a low intensity state with ASM count rate of 0.6 counts s$^{-1}$. We have extracted source spectra from combined source counts of MECS detectors selecting regions of radius 6\arcmin centered on the source and applied time filter for the post egress part. Background spectra was generated from source-free regions in the field of view with the same radius. We have used September 1997 MECS1 response matrices for spectral fitting. Spectral fitting was performed for the 1.5-10.0 keV energy range. After appropriate background subtraction, both 1997 and 2000 spectra were fitted using a single power-law model with a line of sight absorption. The iron emission line is evident at 6.6 keV and is fitted with a Gaussian model. In the low state, the spectrum is harder with a photon index of $\sim$ 0.65 while in high state, the spectrum in contrast is softer with a photon index of $\sim$ 1.18. The iron line equivalent width is also larger in the low state compared to the high state. The hydrogen column density value is comparable in both states.  \\

 We have used the {\sl ASCA} observation of Cen X-3 made during 1993 June 24-25 with the two GIS instruments with a total useful exposure of 39 ks. The light curves from both GIS2 and GIS3 detectors were generated by extracting circular regions of 6{\arcmin} radius around the source with a time resolution of 16s. These light curves were then added together to get the summed light curve. The resulting light curve in terms of orbital phase is shown in Figure 2. The {\sl ASCA} observation was for about half of Cen X-3 orbital period covering the eclipse ingress and part of eclipse egress. A long eclipse transition is obvious during this {\sl ASCA} observation. It is not clear from the {\sl ASCA} light curve if the source is completely out-of-eclipse at the beginning and end of the observation. {\sl ASCA}-GIS spectrum was extracted from the data obtained from individual GIS detectors selecting circular regions of same radius used for light curve extraction from source centered regions. The two GIS spectra were added together using the {\sl addascaspec} tool to obtain the {\sl ASCA} source and background spectra. Auxiliary response file was obtained using the task {\sl ascaarf} and response file gis3v4{\textunderscore}0 is obtained from the {\sl ASCA} database directly. After appropriate background subtraction, the  continuum spectra were fitted in 0.7-10.0 keV energy range using a power-law with Gaussian and line of sight absorption components. The equivalent width of the neutral iron line is large compared to the other observations. \\

 We have used data observed with {\sl XMM} EPIC-PN during 2001 January 27-28 for a total elapsed time of 70 ks in the energy range 0.15-15 keV. The EPIC-PN event files were processed following the standard analysis method of the {\sl XMM} Science Analysis System (SAS) software package. Light curve was obtained by selecting circular extraction region of 30{\arcsec} radius around the source. The observation covers the eclipse egress part. The flux in the egress part increases with a slight nonlinear variation in the intensity. From {\sl RXTE}-ASM data, the source was found to be in an intermediate state during the {\sl XMM} observation with an ASM count rate of 1 counts s$^{-1}$. The source and the background spectra were extracted from circular regions of radius 30\arcsec centered on the source and a source free region respectively and the events were checked for standard filtering before extraction. Response matrices and ancillary response files were generated using the SAS tasks {\sl rmfgen} and {\sl arfgen}. The spectrum was then rebinned so as to have a minimum of 25 counts per spectral channel. The continuum was first fitted with a power-law plus a line of sight absorption. The PN spectrum has very good low energy coverage and it showed  presence of a soft excess. So we added a blackbody component which improved the {$ \chi ^2$} value. The iron line could be resolved into three components and therefore three Gaussian components were included to get the best fitting. \\

Cen X-3 was observed with the ACIS abroad the {\sl Chandra} X-ray Observatory (Weisscopf et al. 2002) on 2007 September 12 for a total exposure of 39 ks. ACIS observation of Cen X-3 covers the egress part of the eclipse. The ACIS event file was analysed using the CIAO v4.0 data analysis software package developed by the {\sl Chandra} X-ray Center (CXC). Light curve was extracted from a circular region of 6{\arcsec} radius with a time resolutions of 100 s. The ASM light curve shows that during the {\sl Chandra} observation the source was in a low state with an average count rate of 0.8 counts s$^{-1}$ and a smooth gradual increase in intensity is found during eclipse egress. To create energy spectrum we extracted data from circular regions of 6\arcsec radius. The source and background spectra were obtained using the standard CIAO tools. The spectrum was extracted using the tool {\sl psextract}. Response files and ancillary files were created using the {\sl mkacisrmf} and {\sl mkarf} tasks. For continuum analysis, we binned the data to have a minimum of 15 counts per bin fitted with a simple power-law modified with photo-electric absorption and a Gaussian component. Figure 3 shows the X-ray spectra of Cen X-3 in high and low intensity states from different observatories along with the best fitted spectral models. Best values for the spectral parameters obtained are shown in Table 2.

\begin{figure}[htbp]
\centering
\includegraphics[width=15 cm, angle=-90]{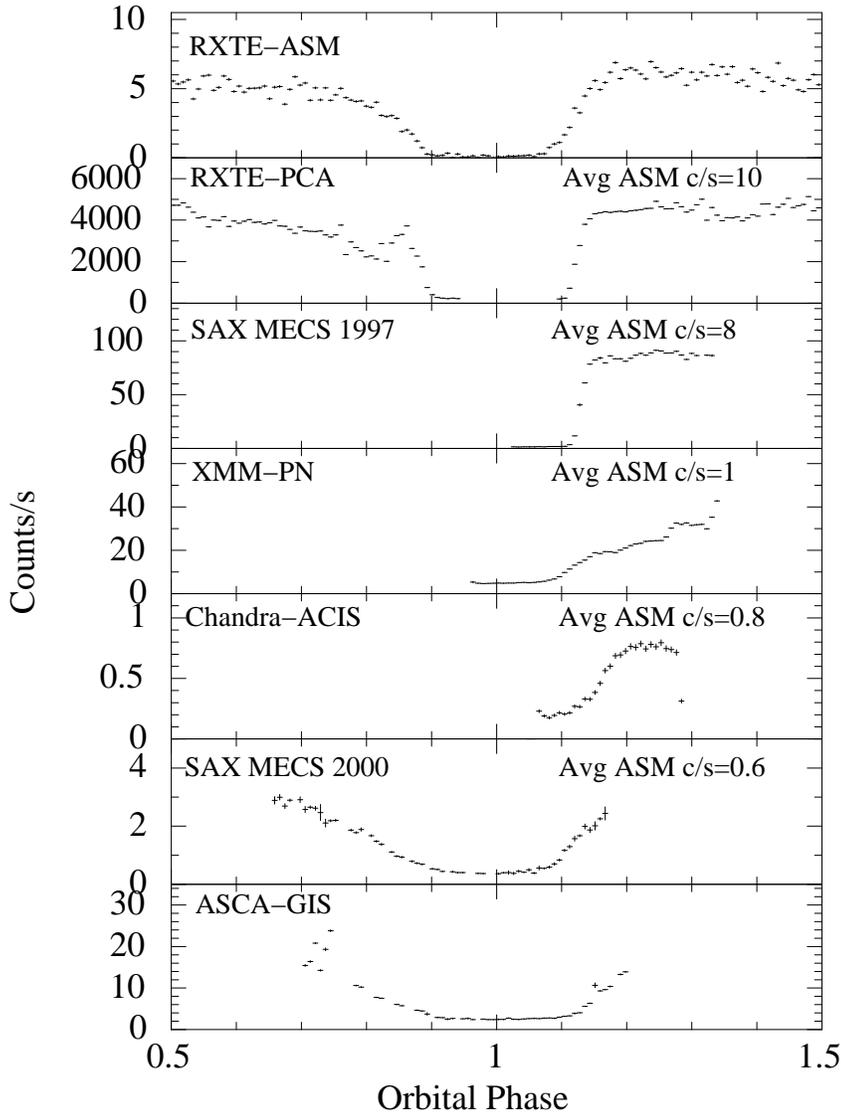}
\caption{The light curves obtained from various missions are shown here.}
\end{figure}

\begin{figure}
\centering
\includegraphics[height=7 cm, angle =-90]{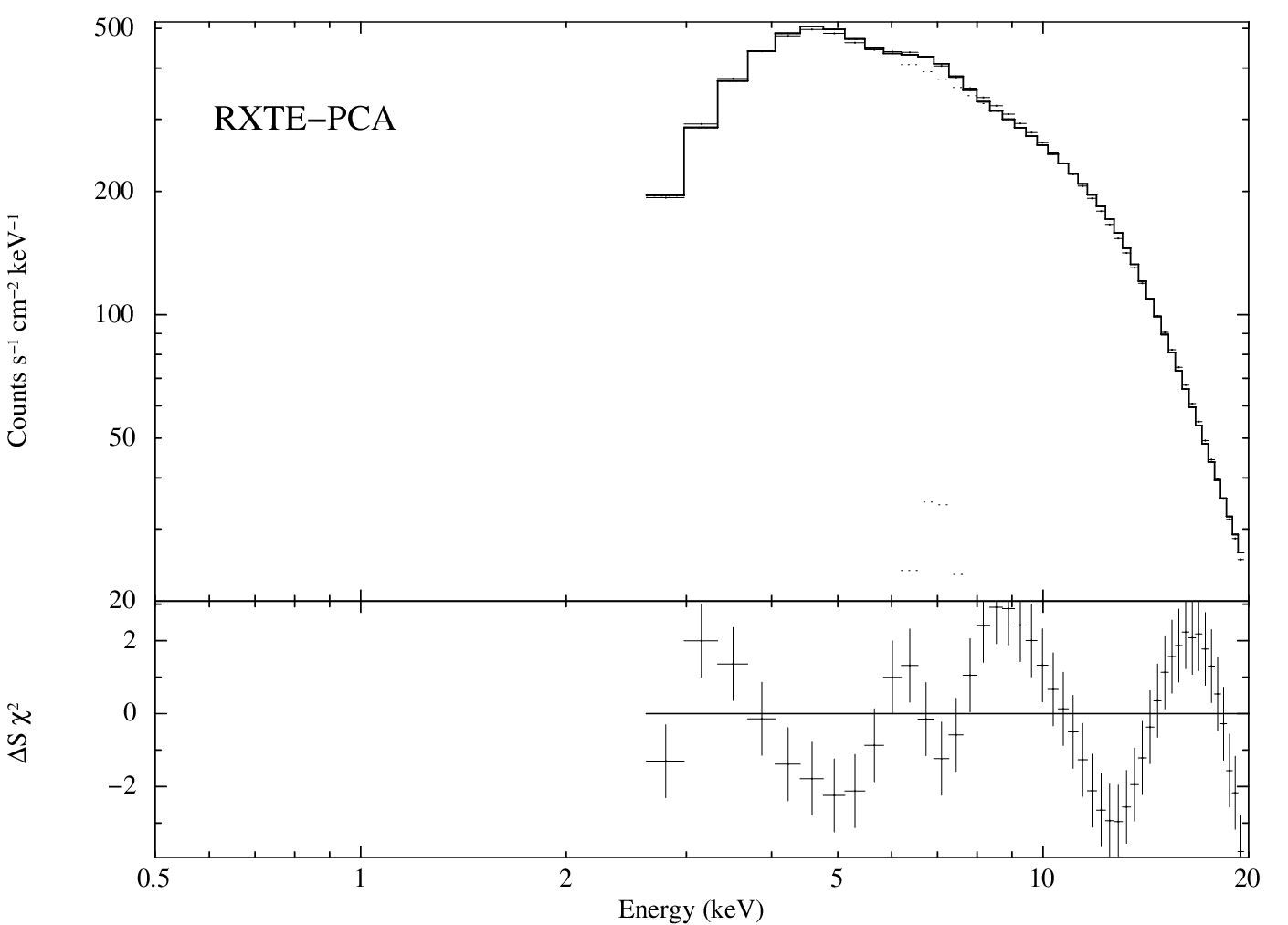}  \includegraphics[height=7 cm, angle =-90]{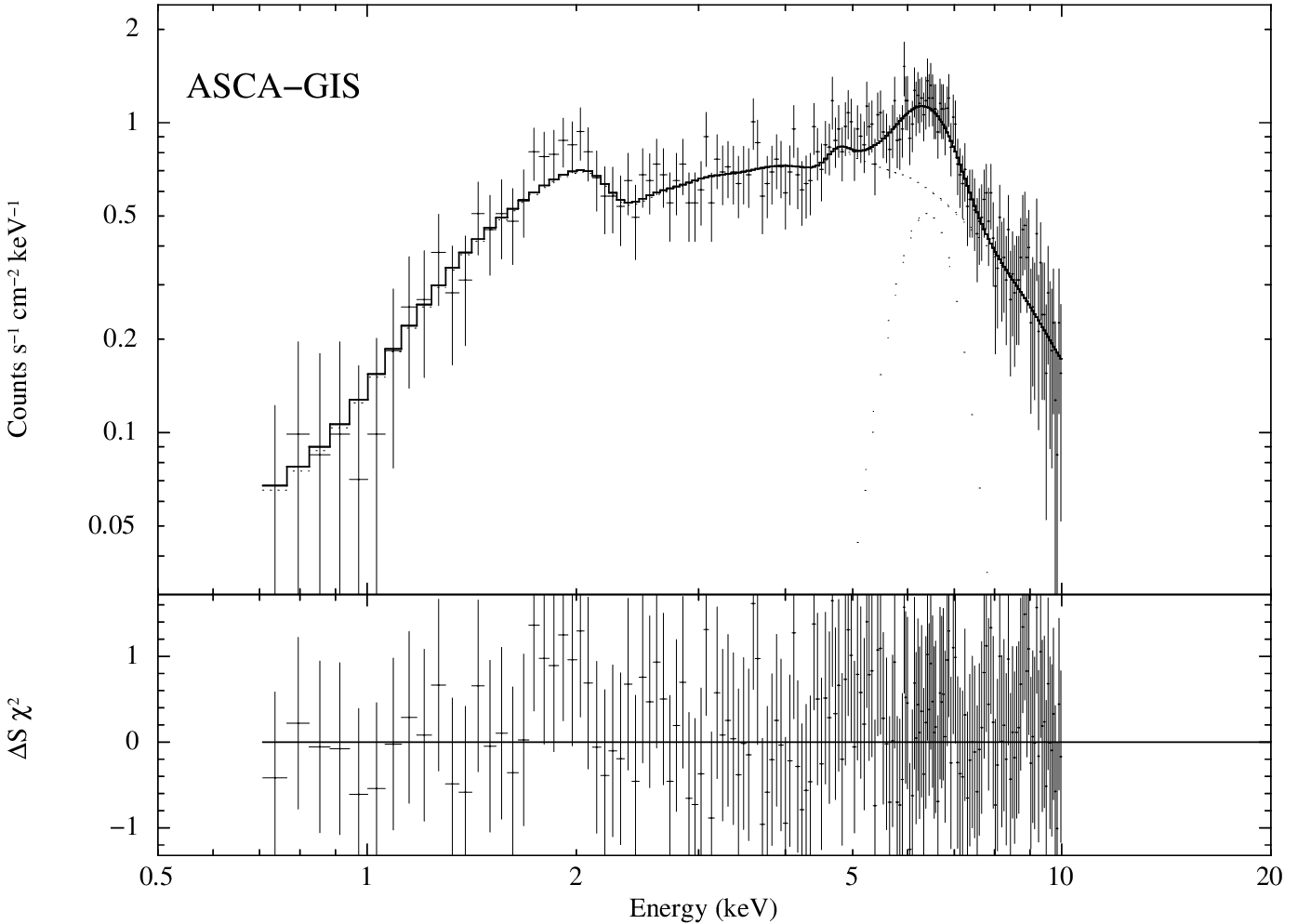}
\includegraphics[height=7 cm, angle =-90]{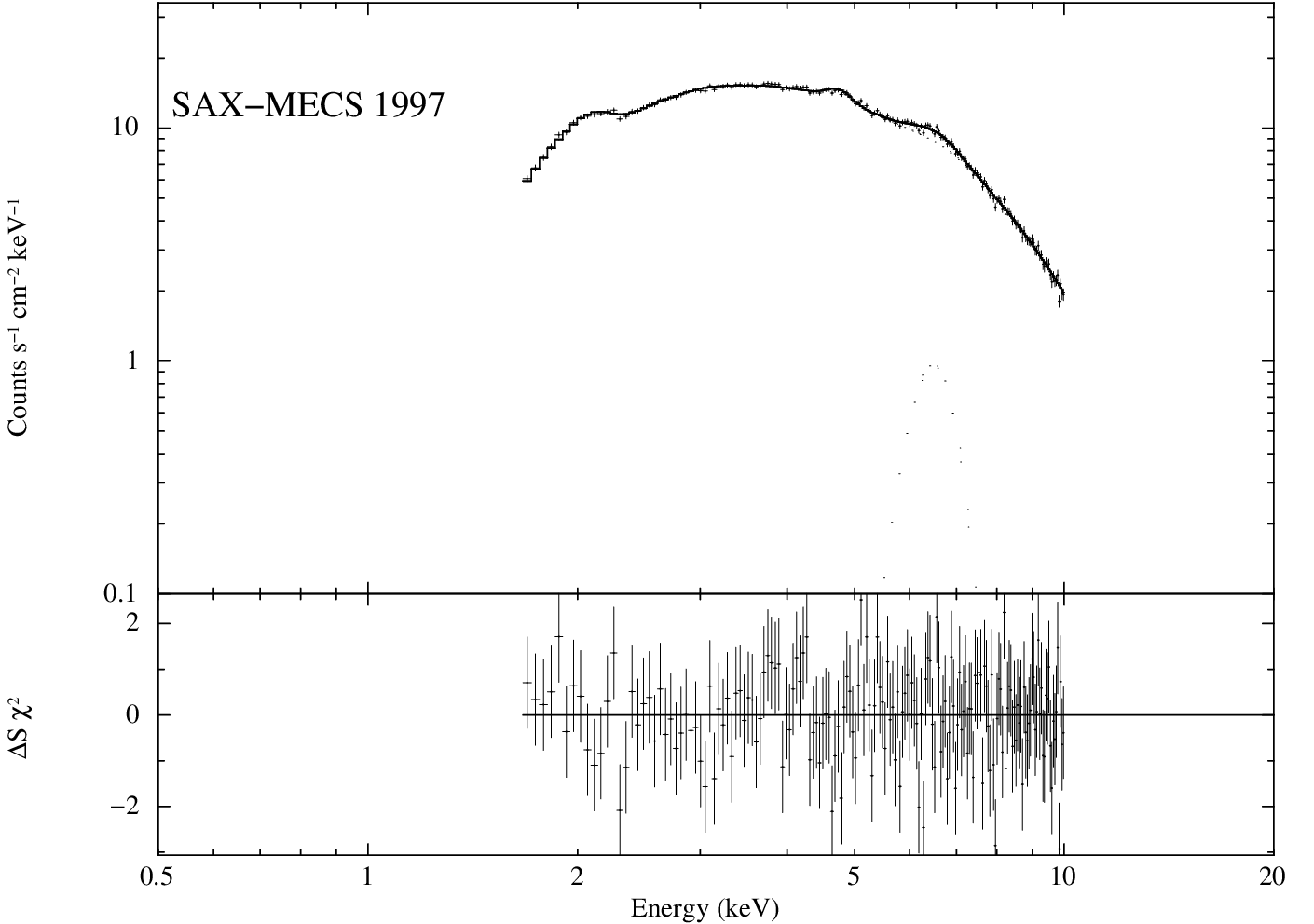}  \includegraphics[height=7 cm, angle =-90]{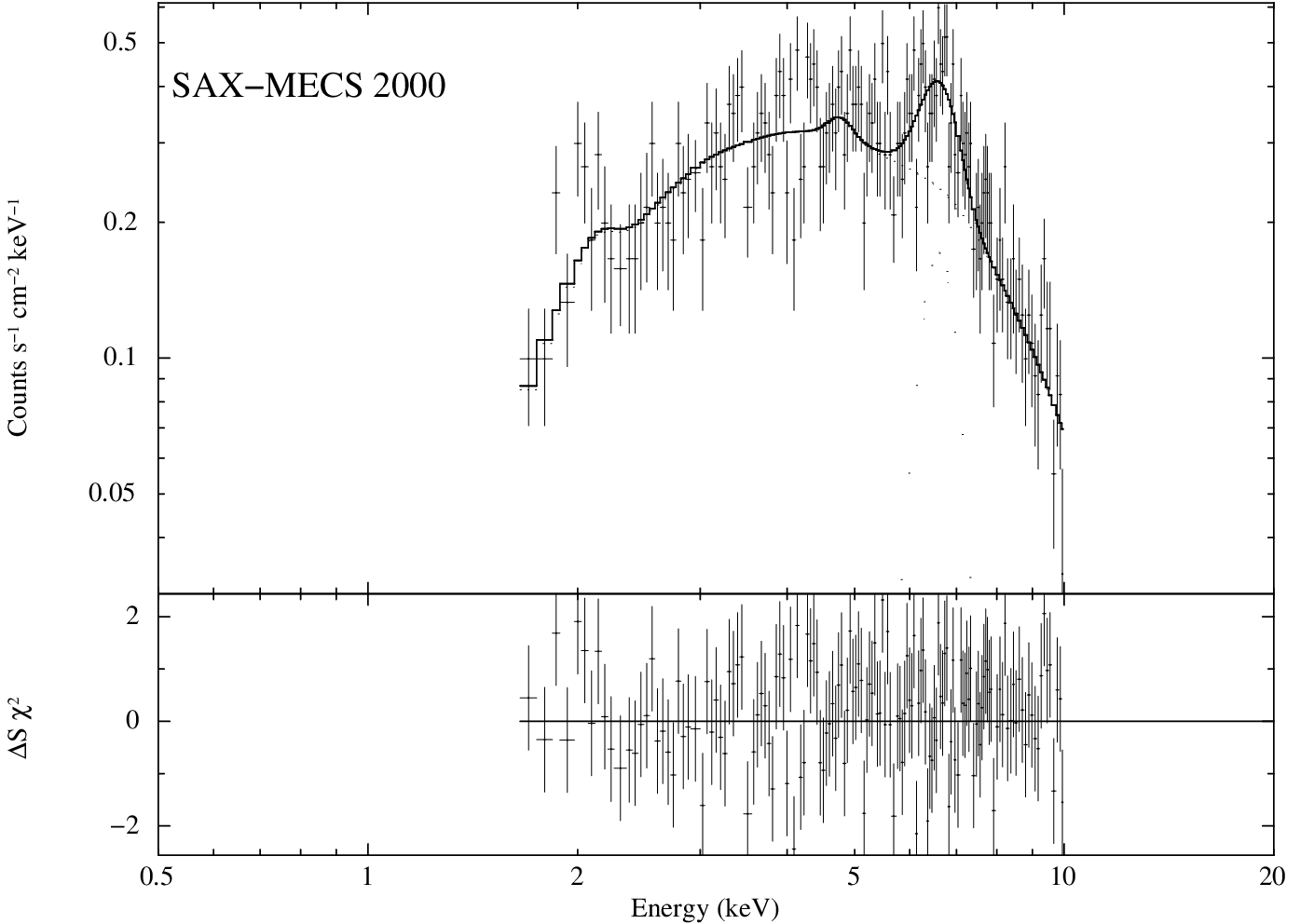} 
\includegraphics[height=7 cm, angle =-90]{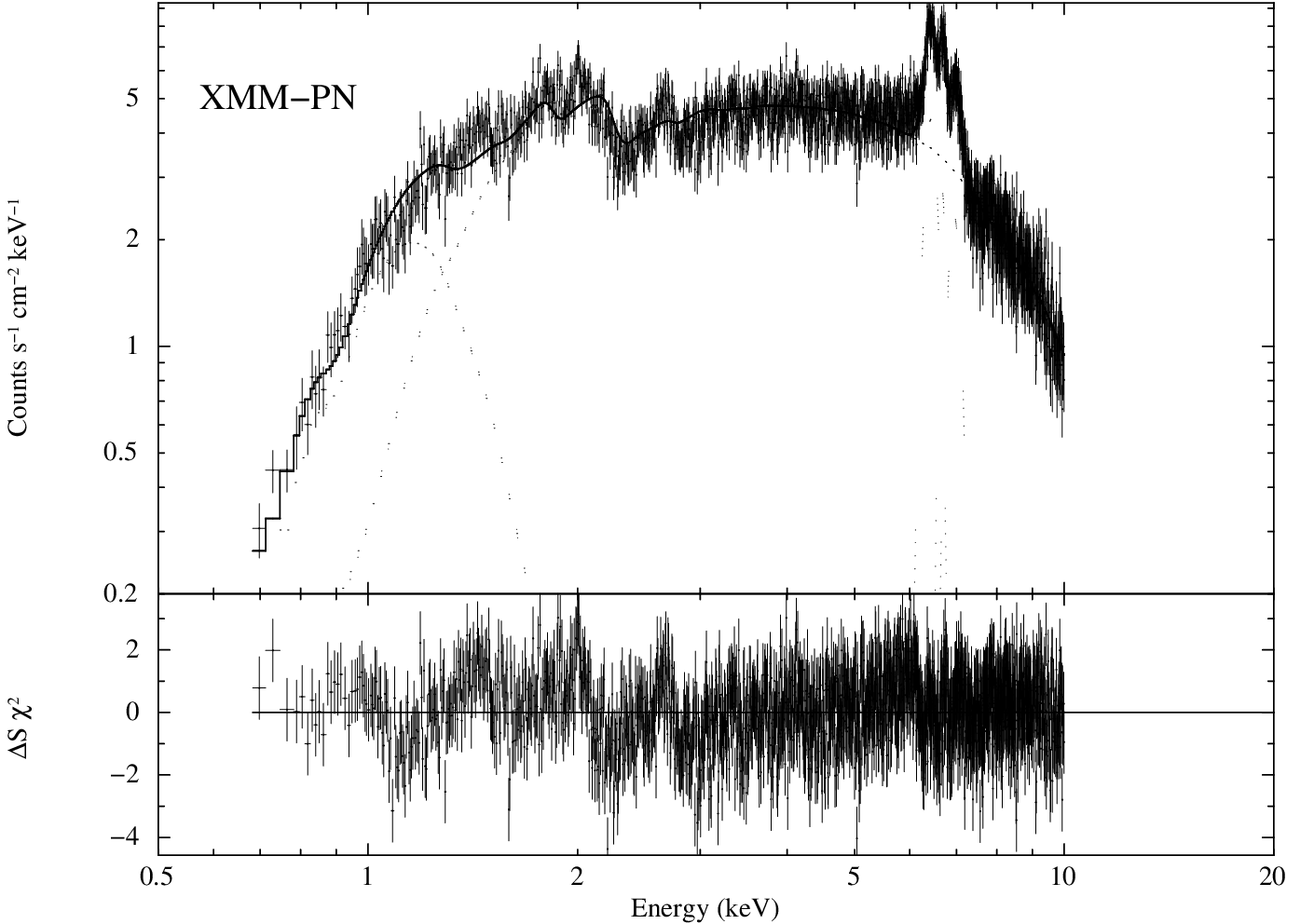}  \includegraphics[height=7 cm, angle =-90]{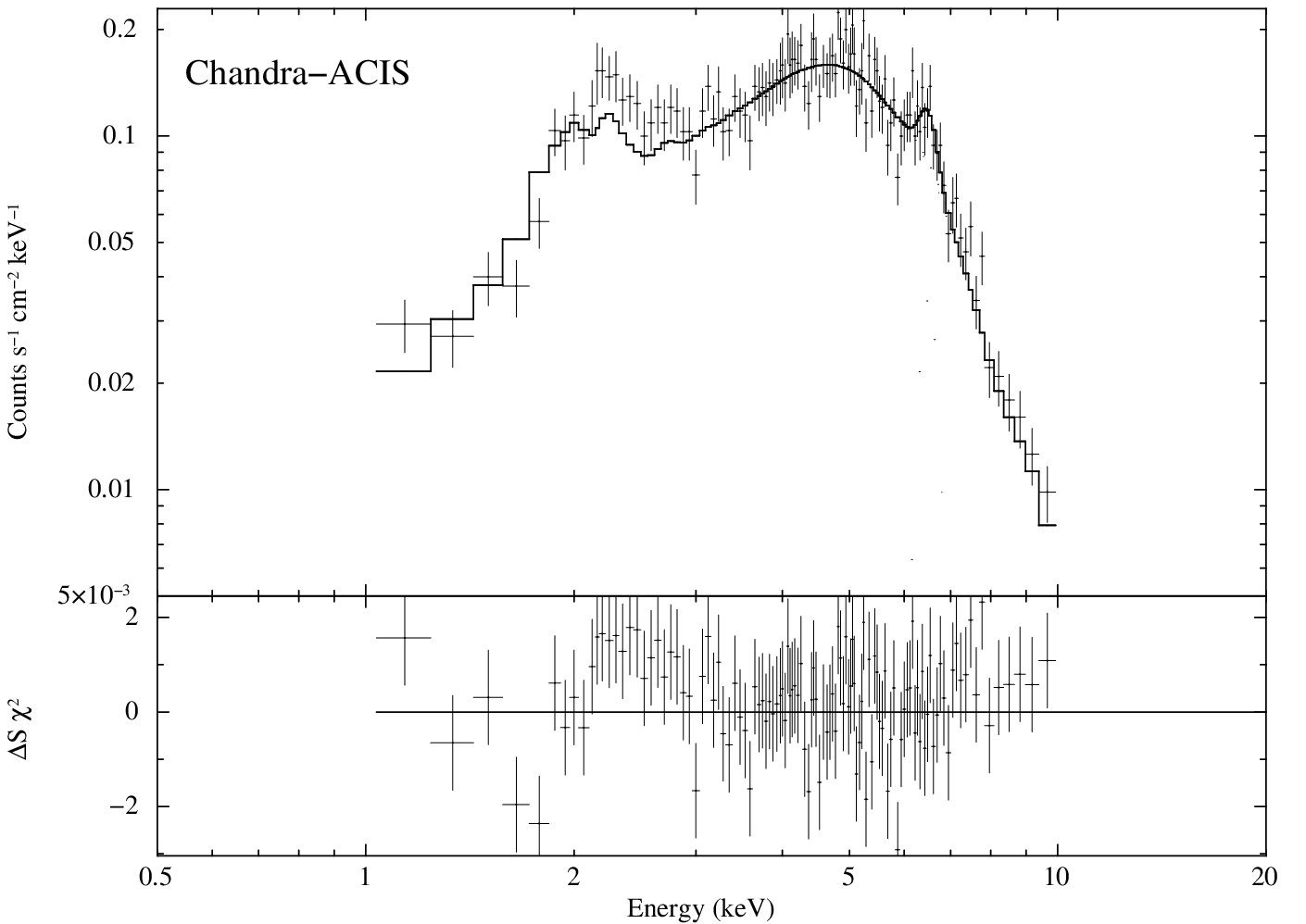}
\caption{Energy spectra of Cen X-3 measured with various missions along with the best-fitted model and their residuals.}
\end{figure}

{
\begin{table}
\caption{Spectral fitting results of Cen X-3}
\begin{center}
\begin{tabular}{l c c c c c c c}
\hline\noalign{\smallskip}
Parameters&{\sl RXTE}- PCA&{\sl SAX}-MECS&{\sl ASCA}-GIS&{\sl XMM} EPIC-PN&{\sl SAX}- MECS &{\sl Chandra}-ACIS \\ 
          &        & 1997   &        &      & 2000    &       \\
\hline\noalign{\smallskip}
%heading
N$_{\rm H}$ $ ^a $&4.3&2.0 $\pm 0.06$&0.5$\pm 0.26$&$1.65_{-0.102}^{+0.032}$&2.2$ \pm0.71$&0.84$\pm 0.26$\\
Photon index&1.36&1.18$\pm0.016$&-0.16$\pm 0.15$&0.44$\pm 0.009$&0.65 $\pm 0.16$&0.30 $\pm 0.125$\\
BB temp (keV)&...&...&...&0.12$\pm 0.005$&...&...\\
BB Norm $ ^b$&...&...&...&6.6$\pm 0.0002$&...&...\\
Fe$_{E}$ (in keV)&6.86&6.6 $\pm 0.066$&6.49 $\pm 0.11$&6.44$\pm 0.007$&6.69 $\pm 0.11$&6.50$\pm 0.10$\\
                                  &...&...            &...            &6.69$\pm 0.007$&...&...\\
                                  &...&...            &...            &6.95&...&...\\
Line Equivalent Width (eV) &136&129&1243&330&750&184\\
              &...    &...&...&80 &...   &...   \\
              &... &...   &...&209&...   &...    \\
Fe line flux $ ^c$&9.4&9.8$\pm0.0017$&10.3$\pm 0.003$&2.18$\pm 0.0001$&2.1$\pm 0.0006$&0.16\\
                             &...&...           &...            &0.77&...&...\\
                             &...&...           &...            &1.27&...&...\\
E$_{cut}$ (in keV)&12.62&...&...&...&...&...\\
E$_{fold}$ (in keV)&10.20&...&...&...&...&...\\
Model flux $ ^d$&5.3&6.0&0.737&0.43&0.23&0.071\\
%\sigma$_{Fe}$ (in eV)
\hline
\end{tabular}
\end{center}
$ ^a ${$10^{22}$ atoms cm$^{-2}$}\\
$ ^b ${$10^{-3}$ photons cm$^{-2}$ s$^{-1}$}\\
$ ^c ${$10^{-3}$ photons cm $^{-2}$ s$^{-1}$}\\
$ ^d ${$10^{-9}$ ergs cm$^{-2}$ s$^{-1}$ for 2-10 keV}

\end{table}
}

\section{Discussion}
 
In the present work, we made an effort to understand the superorbital intensity variations in Cen X-3 using short-term observations with pointed instruments. Compared to the {\sl RXTE}-ASM, the pointed instruments have better statistics in detecting the individual eclipse transitions. In the observations with multiple X-ray observatories reported here we found that in the high state, the light curve shows a sharp ingress and egress with a clear evidence that the observed X-rays are coming directly from the central source. In the low state, the light curve shows a smooth variation in intensity and the eclipse transitions are longer by a significant factor compared to the high state. The observed fluxes in the low state is probably a reprocessed emission from the stellar wind of the companion star. The observations with pointed instruments are thus consistent with long-term observations reported earlier (Raichur \& Paul 2008a) using {\sl RXTE}-ASM light curves.

 {\sl RXTE}-PCA observation and {\sl BeppoSAX} observation, both occurred in the high state of Cen X-3. From the ASM light curve, it is seen that the average count rate during the 1997 {\sl BeppoSAX} observation is 8 counts s$^{-1}$, an intensity of $\sim$ 0.8 compared to the PCA observation. We note here that not only the eclipse transitions last longer in the low state, the bottom four panels of Figure 2 also clearly show that in the low state, the eclipse egress starts earlier by an orbital phase of about 0.02 and also the duration of eclipse is larger compared to the high state. This clearly shows that in the low state, the observed X-rays originate from a much larger region.

In the two {\sl BeppoSAX} observations conducted in high and low states, the out of eclipse X-ray fluxes differ by a factor of 26, whereas the background
subtracted count rates of the eclipse periods in the high and low intensity states differ by a factor of only 4.7. This also indicates that in the low
state, there is an additional scattering medium which scatters some of the source photons towards the observer even when the neutron star is completely
eclipsed. From Table 2, we also see that except for the {\sl Chandra}-ACIS observation, the iron line equivalent width is larger for the low state
observations indicating the presence of significant absorption and reprocessing in the low state. This feature is similar to that seen in Her X-1 and LMC X-4 (Naik \& Paul 2003), two other sources that show strong superorbital X-ray intensity variations.

Even though the observations reported above in different flux states strongly suggested a varying degree of obscuration by the precessing accretion disk,
we cannot rule out the possibility of a part of the variation be due to changes in the mass accretion rate. The long term light curves now being collected
with the highly sensitive {\sl MAXI} Observatory (Matsuoka et al. 2009) will allow measurement of any possible variations in the eclipse durations in several other X-ray binaries with superorbital intensity variations. This will allow us to understand whether precessing accretion disk is a common feature in accreting binaries.

\section*{Acknowledgments} 
  We thank an anonymous referee whose valuable suggestions helped us to improve the content of the paper. This research has made use of data obtained through the High Energy Astrophysics Science Archive Research Center Online Service, provided by the NASA/Goddard Space Flight Center.

\end{document}